\documentclass{article}
\usepackage{amssymb}

\usepackage{graphicx}
\usepackage{amsmath}

\input{tcilatex}

\begin{document}

\title{Gravitational Perturbations of a Radiating Spacetime.}
\author{Manasse R. Mbonye \\
\textit{Physics Department, University of Michigan, Ann Arbor, Michigan 48109%
} \and Ronald L. Mallett \\
\textit{Department of Physics, University of Connecticut, Storrs,
Connecticut, 06269}}
\maketitle

\begin{abstract}
This paper discusses the problem of gravitational perturbations of radiating
spacetimes. We lay out the theoretical framework for describing the
interaction of external gravitational fields with a radiating spacetime.
This is done by deriving the field perturbation equations for a radiating
metric. The equations are then specialized to a Vaidya spacetime. For the
Hiscock ansatz of a linear mass model of a radiating blackhole the equations
are found separable. Further, the resulting ordinary differential equations
are found to admit analytic solutions. We obtain the solutions and discuss
their characteristics.
\end{abstract}

\section{INTRODUCTION.}

The study of gravitational perturbations can be traced back to the famous
Einstein-Infeld-Hoffman paper of 1938 \cite{c1} which pioneered the
treatment of the two body problem in general relativity. In 1957 Regge and
Wheeler \cite{c2} addressed the problem of the stability of a Schwarzschild
black hole. Later, in his study of perturbations of a rotating black hole ( 
\cite{c3} and later papers), Teukolsky was able to put the discipline on a
stronger footing. However, little progress has been made at extending this
success to cover the radiating cases. The problem of perturbing a radiating
spacetime with integral spin fields has not received the attention it
deserves. This, despite the fact that most astrophysical objects radiate.
From regular stars to supernovae, from quasars to primordial black holes one
finds that the inhabitants of our universe are generally non-static.

In the present paper, we develop a framework for discussing the problem of
how external gravitational fields may interact with radiating spacetimes.
This is done by deriving the field perturbation equations. It is found that
two such equations are sufficient to describe all the non-trivial features
of the perturbing gravitational field. We find that one of these equations
decouples completely and is homogeneous in one of the field components. The
equation for the other field component contains, in its source terms,
several perturbed and therefore undetermined quantities. Using a systematic
approach we are able to determine all these quantities completely, in terms
of the former field component. The result is that all the perturbations are
described by only two field components which satisfy two partial
differential equations. The equations are then specialized to a Vaidya
spacetime. For a particular model of a radiating black hole the equations
are found separable. Interestingly, the resulting ordinary differential
equations are found to admit analytic solutions. We obtain these solutions
and discuss their characteristics.

The mathematical framework used in this paper is the null tetrad formalism
of Newman and Penrose (hereafter NP formalism) \cite{c8}. In section\textit{%
\ }II we give a brief description of the background geometry, the radiating
spacetime of Vaidya. In section III we derive the perturbation field
equations for a general non-vacuum type D spacetime and adapt these
equations to the Vaidya spacetime. In section IV we calculate the perturbed
quantities in the source terms to arrive at the final working field
equations. It is demonstrated in section V that these equations are
separable for the Hiscock linear model \cite{c14.1} of a radiating black
hole. In section VI we obtain and discuss some of the solutions and we
conclude the discussion in section VII..

\section{THE\ VAIDYA\ SPACETIME.}

\subsection{\textbf{The Metric}}

In this analysis we perturb the Vaidya spacetime \cite{c11} with incoming
external gravitational fields. The Vaidya geometry, the simplest of the
radiating spacetimes, is non-rotating and spherically symmetric. The
energy-momentum tensor

\begin{equation*}
T_{\mu \nu }=\rho k_{\mu }k_{\nu }, 
\end{equation*}
describes a null fluid, $\left( k_{\mu }k^{\mu }=0\right) $, of density $%
\rho $ with radial flow, $k^{2}=k^{3}=0$. Using this energy-momentum tensor
to solve the Einstein field equations one obtains \cite{c12} a line element,
in retarded coordinates, given by 
\begin{equation}
ds^{2}=\left[ 1-\frac{2m\left( u\right) }{r}\right] du^{2}+2dudr-r^{2}d%
\Omega ^{2}.  \label{e1.4}
\end{equation}
\newline

Here, $u$ is the retarded time coordinate and $m(u)$, the mass, is a
monotonically decreasing function of $u$.

It is convenient to introduce a null tetrad basis $z_{m\mu }=\left( l_{\mu
},n_{\mu },m_{\mu },\bar{m}_{\mu }\right) $ at every point in this
spacetime. The metric tensor $g_{\mu \nu }$ then becomes 
\begin{equation}
g_{\mu \nu }=z_{m\mu }z_{n\nu }\eta ^{mn}=\ l_{\mu }n_{\nu }+n_{\mu }l_{\nu
}-m_{\mu }\bar{m}_{\nu }-\bar{m}_{\mu }m_{\nu },  \label{e1.51}
\end{equation}
where $\eta ^{mn}$ \thinspace is the flat spacetime metric. Following
Carmeli and Kaye \cite{c10} we choose the covariant form of the null tetrad
basis as

\begin{equation*}
l_\mu =\delta _\mu ^{\,\,0}, 
\end{equation*}

\begin{equation*}
\;n_\mu =\frac 12\left[ 1-\frac{2m\left( u\right) }r\right] \delta _\mu
^{\;0}+\delta _\mu ^{\;1}, 
\end{equation*}

\begin{equation}
m_\mu =-\frac r{\sqrt{2}}\left[ \delta _\mu ^{\;2}+i\;\sin \,\theta \delta
_\mu ^{\;3}\right] ,  \label{e1.5}
\end{equation}

\begin{equation*}
\bar{m}_{\mu }=-\frac{r}{\sqrt{2}}\left[ \delta _{\mu }^{\;2}-i\;\sin
\,\theta \delta _{\mu }^{\;3}\right] 
\end{equation*}
The contravariant vectors, $z_{m}$, considered as tangent vectors, define
the directional derivatives as 
\begin{equation}
\vec{z}_{m}=z_{m}^{\;\mu }\nabla _{\mu }  \label{e1.6}
\end{equation}

In the Vaidya spacetime these directional derivatives are given (from
equation \ref{e1.5}) by

\begin{eqnarray}
D &=&l^{\mu }\nabla _{\mu }=\frac{\partial }{\partial \,r},  \notag \\
\Delta &=&n^{\mu }\nabla _{\mu }=\frac{\partial }{\partial \,u}-\frac{1}{2}%
\left[ 1-\frac{2m\left( u\right) }{r}\right] \frac{\partial }{\partial \,r},
\notag \\
\delta &=&m^{\mu }\nabla _{\mu }=\sqrt{2}\,r\left[ \frac{\partial }{\partial
\,\theta }+i\;\csc \,\theta \frac{\partial }{\partial \,\varphi }\right] , 
\notag \\
\bar{\delta} &=&\bar{m}^{\mu }\nabla _{\mu }=\sqrt{2}\,r\left[ \frac{%
\partial }{\partial \,\theta }-i\;\csc \,\theta \frac{\partial }{\partial
\,\varphi }\right] .  \label{e1.7}
\end{eqnarray}

One finds that the only surviving spin (Ricci rotation) coefficients \cite
{c10} of the Vaidya spacetime are

\begin{eqnarray}
\rho &=&-\left( \frac{1}{r}\right) ,\;\;\alpha =-\frac{1}{2\sqrt{2}\;r}%
\;\cot \,\theta ,\;\;\beta =-\alpha ,  \notag \\
\mu &=&-\frac{1}{2r}\left[ 1-\frac{2m\left( u\right) }{r}\right] ,\
\;and\;\;\gamma =\frac{m\left( u\right) }{2\,r^{2}}  \label{e1.71}
\end{eqnarray}

The only surviving tetrad component of Ricci tensor \cite{c10} is

\begin{equation}
\Phi _{22}=\frac{-\dot m\left( u\right) }{r^2},  \label{e1.9}
\end{equation}
where $\dot m=\frac{dm}{du}$. With this component, it is easily shown that

\begin{equation}
T_{\mu \nu }=2\Phi _{22}l_\mu l_\nu =-\left[ 2\frac{\dot m\left( u\right) }{%
r^2}\right] l_\mu l_\nu .  \label{e1.10}
\end{equation}

Further, the only non-vanishing tetrad component of the of the Weyl tensor is

\begin{equation}
\Psi _2=\frac{-m\left( u\right) }{r^3}.  \label{e1.11}
\end{equation}

The Vaidya spacetime is then \cite{c8} said to be Petrov type D with
repeated principal null vectors $l^{\mu }$ and $n^{\mu }$. The three optical
scalars are found, to be $\sigma =\omega =0,\;\;\theta =-\frac{1}{r}$. The
metric contains two shear-free, twistless and diverging geodetic null
congruencies.

\ \ \ 

\section{The Perturbed \textbf{Field Equations}}

\subsection{\textbf{The Type D Spacetime}.}

In this section we develop the gravitational field perturbation equations
for a general Petrov type D spacetime. We start with the Newman-Penrose (NP)
equations. The full set of the NP equations can be found in several
publications, for example \cite{c10}. We only mention here that the set is
made up of first order differential equations which, in the NP formalism,
replace the Einstein field equations. The equations link together the null
tetrad basis, the spin (Ricci rotation) coefficients, the Weyl tensor, the
Ricci tensor and the scalar curvature. In using this formalism to do
perturbation analysis one first specifies the perturbations of the geometry.
Here we shall write the tetrad of the perturbed spacetime as

\begin{equation}
\mathbf{l}=\mathbf{l}^b+\mathbf{l}^p,\;\;\mathbf{n=n}^b+\mathbf{n}^p,\;\;%
\mathbf{m=m}^b\mathbf{+m}^p,\;\;\mathbf{\bar m=\bar m}^b\mathbf{+\bar m}^p.
\label{e2.1}
\end{equation}
where the superscripts $b$ and $p$ refer to the background and the
perturbing quantities, respectively. Since all the other field quantities
are expressible in terms of the tetrad \cite{c9} their perturbed forms can
be written down. For example: $\Psi _{a\,=\,\left( 0,1,2,3,4\right) }=\Psi
_a^{\;b}+\Psi _a^{\;p}$.

We shall, in general, assume that the perturbations in the basis vectors are
sufficiently small so that only their first order contributions are
significant. The field equations are, then, first order in the perturbing
fields, linearized about the background quantities.

In type D spacetimes one can choose the tetrad vectors (see section II) $%
\mathbf{l}$ and $\mathbf{n}$ so that certain spin coefficients and certain
Weyl scar components

\begin{eqnarray}
\kappa ^{b} &=&\sigma ^{b}=\nu ^{b}=\lambda ^{b}=0,  \notag \\
\Psi _{0}^{\;b} &=&\Psi _{1}^{\;b}=\Psi _{3}^{\;b}=\Psi _{4}^{\;b}=0.
\label{e2.2}
\end{eqnarray}
We start our analysis from three of the NP \cite{c10} field equations. From
the Bianchi identities we consider the two equations,

\begin{equation*}
\bar \delta \Psi _3-D\Psi _4+\bar \delta \Phi _{21}-\Delta \Phi
_{20}=3\lambda \Psi _2-2\left( \alpha +2\pi \right) \Psi _3+\left( 4\epsilon
-\rho \right) \Psi _4 
\end{equation*}

\begin{equation}
-2\nu \Phi _{10}+2\lambda \Phi _{11}+\left( 2\gamma -2\bar \gamma +\bar \mu
\right) \Phi _{20}+2\left( \bar \tau -\alpha \right) \Phi _{21}-\bar \sigma
\Phi _{22},  \label{e2.3}
\end{equation}
\ \ and

\begin{equation*}
\Delta \Psi _{3}-\delta \Psi _{4}+\bar{\delta}\Phi _{22}-\Delta \Phi
_{21}=3\nu \Psi _{2}-2\left( \gamma +2\mu \right) \Psi _{3}+\left( 4\beta
-\tau \right) \Psi 
\end{equation*}

\begin{equation}
-2\nu \Phi _{11}-\bar \nu \Phi _{20}+2\lambda \Phi _{12}+2\left( \gamma
+\bar \mu \right) \Phi _{21}+\left( \bar \tau -2\bar \beta -2\alpha \right)
\Phi _{22},  \label{e2.4}
\end{equation}
from the spin coefficient system of equations we take

\begin{equation}
\Delta \lambda -\bar \delta \nu =-\left( \mu +\bar \mu \right) \lambda
-\left( 3\gamma -\bar \gamma \right) \lambda +\left( 3\alpha +\bar \beta
+\pi -\tau \right) \nu -\Psi _4.  \label{e2.41}
\end{equation}
We have complete knowledge of the geometry of the background spacetime, so
we write down the equations only in those terms that make first order
contributions to the perturbed field quantities. Making use of equations \ref
{e2.2} on equations \ref{e2.3}, \ref{e2.4} and \ref{e2.41}, respectively, it
is seen that for a perturbed Petrov type D spacetime

\begin{equation*}
-3\lambda ^p\Psi _2+\left( \bar \delta +2\alpha +4\pi \right) \Psi
_3^{\;p}-\left( D+4\epsilon -\rho \right) \Psi _4^{\;p} 
\end{equation*}

\begin{equation}
=\left( \Delta +2\gamma -2\bar \gamma +\mu \right) \Phi _{20}^{\;\,p}-\left(
2\alpha -2\bar \tau +\bar \delta \right) \Phi _{21}^{\;\,p}-\bar \sigma
^p\Phi _{22},  \label{e2.5}
\end{equation}

\ \ 

\begin{equation*}
-3\nu ^p\Psi _2+\left( \Delta +2\gamma +4\mu \right) \Psi _3^{\;p}-\left(
4\beta -\tau +\delta \right) \Psi _4^{\;p}=\left( 2\gamma +2\bar \mu +\Delta
\right) \Phi _{21}^{\;\,p} 
\end{equation*}

\begin{equation}
+\left( -\bar \delta +\bar \tau -2\bar \beta -2\alpha \right) ^p\Phi
_{22}+\left( -\bar \delta +\bar \tau -2\bar \beta -2\alpha \right) \Phi
_{22}^{\;\,p},  \label{e2.6}
\end{equation}
and

\begin{equation}
\left( \Delta +3\gamma -\bar \gamma +\mu +\bar \mu \right) \lambda ^p-\left(
\bar \delta +3\alpha +\bar \beta +\pi -\bar \tau \right) \nu ^p+\Psi
_4^{\;p}=0.  \label{e2.7}
\end{equation}

We note that the perturbed equations are coupled both in the Weyl tensor
components and the Ricci tensor components. They also contain unknown spin
coefficients and directional derivatives. In attempting to decouple the
equations above we use an approach akin to Teukolsky's \cite{c3}. After some
algebra we find, on eliminating $\Psi _{3}$ between the equations \ref{e2.5}
and \ref{e2.6}, we are left with

\begin{equation*}
\left[ \left( \Delta +3\gamma -\bar \gamma +4\mu +\bar \mu \right) \left(
D+4\epsilon -\rho \right) \right] \Psi _4^{\;p} 
\end{equation*}

\begin{equation}
-\left[ \left( \bar{\delta}+3\alpha +\bar{\beta}+4\pi -\bar{\tau}\right)
\left( 4\beta -\tau +\delta \right) -3\Psi _{2}\right] \Psi _{4}^{\;p}=Q_{4},
\label{e2.8}
\end{equation}

Now, under the interchange \textbf{l\thinspace }$\rightleftarrows \,$\textbf{%
n}, \textbf{m\thinspace }$\rightleftarrows \,$\textbf{\={m}}, the full set
of the NP equations is invariant \cite{c13}. This symmetry is not destroyed
by the choice of \textbf{l }and \textbf{n} which gave equations \ref{e2.2}.
One finds under this interchange that \cite{c9}

\begin{equation*}
\Psi _0\rightleftarrows \Psi _4^{*},\;\Psi _3\rightleftarrows \Psi
_1^{*},\;\Psi _2\rightleftarrows \Psi _2^{*}, 
\end{equation*}

\begin{equation}
\kappa \rightleftarrows -\bar{\nu},\;\rho \rightleftarrows -\bar{\mu}%
,\;\sigma \rightleftarrows -\bar{\lambda},\;\alpha \rightleftarrows -\bar{%
\beta},\;\epsilon \rightleftarrows -\bar{\gamma}\;and\;\pi \rightleftarrows -%
\bar{\tau},  \label{e2.12}
\end{equation}
Applying this to equation \ref{e2.8} we obtain the following equation for $%
\Psi _{0}^{\;p}$

\begin{equation*}
\left( D-3\epsilon -\bar \epsilon -4\rho -\bar \rho \right) \left( \delta
-3\beta -\bar \alpha -4\tau +\bar \pi \right) \left( \bar \delta -4\alpha
+\pi \right) \Psi _0^{\;p} 
\end{equation*}

\begin{equation}
-\left( D-3\epsilon -\bar{\epsilon}-4\rho -\bar{\rho}\right) \left[ \left(
\Delta -4\gamma +\mu \right) -3\Psi _{2}\right] \Psi _{0}^{\;p}=Q_{0}.
\label{e2.13}
\end{equation}
Here, the source term $Q_{0}$ is given by

\begin{equation*}
Q_0=\left( \delta -3\beta -\bar \alpha -4\tau +\bar \pi \right) \left(
D-2\bar \rho -2\epsilon \right) \Phi _{02}^{\;\,p} 
\end{equation*}

\begin{equation*}
-\left( \delta -3\beta -\bar \alpha -4\tau +\bar \pi \right) \left( \delta
+\bar \pi -2\bar \alpha -2\beta \right) \Phi _{00}^{\;\,p} 
\end{equation*}

\begin{equation*}
+\left( \delta -3\beta -\bar \alpha -4\tau +\bar \pi \right) \left( \delta
+\bar \pi -2\bar \alpha -2\beta \right) ^p\Phi _{00}^{\;} 
\end{equation*}

\begin{equation*}
+\left( D-3\epsilon +\bar \epsilon -4\rho -\bar \rho \right) \left( \delta
-2\beta +2\bar \pi \right) \Phi _{02}^{\;\,p} 
\end{equation*}

\begin{equation}
-\left( D-2\epsilon +2\bar \epsilon -\rho \right) \Phi _{22}^{\;\,p}+\left(
D-3\epsilon +\bar \epsilon -4\rho -\bar \rho \right) \sigma ^p\left( \Delta
-\mu \right) \Phi _{00}.  \label{e2.14}
\end{equation}

Equations \ref{e2.8} and \ref{e2.13} describe the gravitational field
perturbation equations for a general type D spacetime with sources.

\subsection{The Perturbed Vaidya Spacetime.}

We shall now adapt the equations above to the problem of perturbing a Vaidya
spacetime with gravitational fields. The important features to deal with
here are the perturbations in the source terms $Q_{0}$ and $Q_{4}$ of
equations \ref{e2.8} and \ref{e2.13}.

The energy-momentum tensor associated with the radiation in the Vaidya
spacetime is usually interpreted via geometrical optics. Carmeli and Kaye 
\cite{c10} have, in fact, shown that the associated radiation field which
has a monopole structure can not be identified as a source-free
electromagnetic field. Therefore in addressing the problem of perturbing
this radiation, one can only discuss direct perturbations on the
energy-momentum tensor. Now, this energy-momentum tensor is quadratic in the
metric \cite{c15}. It follows from these considerations, then, that the
lowest order perturbations in the scalar components $\Phi _{mn}$ will be
quadratic in the metric perturbations. However, we assumed from the
beginning that the perturbations in the basis vectors are sufficiently small
so that only their first order contributions are significant. In our linear
theory we shall, therefore, disregard these perturbations in the scalar
components $\Phi _{mn}$ contributions. Nevertheless, we still have various
perturbed spin coefficients and directional derivatives to deal with in the
source terms.

It is recalled (see section II) that the only quantities which survive in
the Vaidya spacetime background are: the spin coefficients $\alpha $, $\beta 
$, $\gamma $, $\rho $, $\mu $, the Weyl tensor component $\Psi _2$ and the
Ricci tensor component $\Phi _{22}$. Moreover, all these spin coefficients
are real and $\,\alpha =-\beta $. On applying the observations made above to
the field perturbation equations \ref{e2.8} and \ref{e2.13} we find that

\begin{equation}
\left[ \left( \Delta +2\gamma +5\mu \right) \left( D-\rho \right) -\left(
\bar \delta -2\beta \right) \left( \delta +4\beta \right) -3\Psi _2\right]
\Psi _4^{\;p}=Q_4^{\prime },  \label{C35}
\end{equation}

\begin{equation}
\left[ \left( D-5\rho \right) \left( \Delta -4\gamma +\mu \right) -\left(
\delta -2\beta \right) \left( \bar{\delta}+4\beta \right) -3\Psi _{2}\right]
\Psi _{0}^{\;p}=0,  \label{C36}
\end{equation}
where the source terms are now given by 
\begin{equation}
Q_{4}^{\prime }=\left[ \left( \Delta +2\gamma +5\mu \right) \bar{\sigma}%
^{p}-\left( \bar{\delta}-2\beta \right) \left( \bar{\delta}^{p}-\bar{\tau}%
^{p}+2\bar{\beta}^{p}+2\alpha ^{p}\right) +\lambda ^{p}\left( D+\rho \right) %
\right] \Phi _{22},  \label{C37}
\end{equation}
and clearly 
\begin{equation}
Q_{0}^{\prime }=0.  \label{C38}
\end{equation}
Note that in equation \ref{C37} the derivative operators $\Delta $ and $D$
must act on $\Phi _{22}$.

We now have two equations, \ref{C35} and \ref{C36} which describe the two
Weyl tensor components $\Psi _{0}^{\;p}$ and $\Psi _{4}^{\;p}$. It can be
shown that these two components are sufficient to describe all the
non-trivial features of the perturbing fields. The proof of this sufficiency
is achieved by showing \cite{c3} that only $\Psi _{0}^{\;p}$ and $\Psi
_{4}^{\;p}$ are invariant under gauge transformations and infinitesimal
tetrad rotations. This invariance, in turn, makes them completely measurable
physical quantities.

The source terms $Q_{4}^{\prime }$ in equation \ref{C35} still contain (see
equation \ref{C37}) perturbations in as many as seven different quantities.
In the next section we shall derive equations that describe these unknowns.
In doing so, we shall discover that those perturbed quantities that do not
vanish are all completely describable in terms of the one of the two field
components.

\section{The perturbed quantities in the source terms.}

The aim of this section is to calculate, (in terms of the known background
quantities and the fields $\Psi _{0}^{\;p}$ and $\Psi _{4}^{\;p}$) the
perturbed quantities, $\bar{\Phi}_{12}^{p},$ $\alpha ^{p},\;\beta
^{p},\;\tau ^{p},\;\sigma ^{p},\;\lambda ^{p}$ and $\delta ^{p}$ which
appear in the source terms $Q_{4}^{\prime }$ of equation \ref{C35}, (see
also equation \ref{C37}). It is worthwhile to point out that the system of
equations we have constructed in the previous section must be consistent
with the freedom we have in the choice of both the tetrad frame and the
coordinates. In particular \cite{c9}, from the 6-parameter group of
homogeneous Lorentz transformations, we have six degrees of freedom to make
infinitesimal rotations of the local tetrad-frame. Further, we have four
degrees of freedom to make infinitesimal coordinate transformations. Thus,
altogether we have a total of ten degrees of gauge freedom. We are free to
exercise these available degrees of freedom as convenience and occasion may
dictate.

As has been shown \cite{c3}, in a linear perturbation theory, $\Psi
_{0}^{\;p}$ and $\Psi _{4}^{\;p}$ are gauge invariant while $\Psi _{1}^{\;p}$%
, $\Psi _{2}^{\;p}$ and $\Psi _{3}^{\;p}$ are not. This means that we can
subject the tetrad null-basis to an infinitesimal rotation in which $\Psi
_{1}^{\;p}$ and $\Psi _{3}^{\;p}$ vanish without affecting $\Psi _{0}^{\;p}$
and $\Psi _{4}^{\;p}$. In making this choice of gauge, we have used up only
four, out of ten, degrees of freedom. It can be shown (see for example \cite
{c9} Chapter 1, equations 342 and 346) that in linearized type 1 rotations
both $\Psi _{1}^{\,\,p}$ and $\bar{\Phi}_{12}^{p}$ vanish while in type 2
rotations both do not. It follows, then, that a gauge in which both $\Psi
_{1}^{\;p}$, $\Psi _{3}^{\;p}$ and $\bar{\Phi}_{12}^{p}$ vanish can be
chosen. Further, in this gauge, $\Psi _{2}^{\;p}$ clearly vanishes. Under
these circumstances, the NP equations \cite{c9} show that the linearized
Bianchi identities take on a simpler form. For our purposes, the equations
we need to consider from this set are:

\begin{equation}
\Delta \Psi _0^{\;p}=\left( 4\gamma -\mu \right) \Psi _0^{\;p}+3\sigma
^p\Psi _2,  \label{e4.1}
\end{equation}

\begin{equation}
-3\delta ^p\Psi _2=-9\tau ^p\Psi _2-2\kappa ^p\Phi _{22},  \label{e4.2}
\end{equation}

\begin{equation}
-D\Psi _4^{\;p}=3\lambda ^p\Psi _2-\rho \Psi _4^{\;p}-\bar \sigma ^p\Phi
_{22},  \label{e4.4}
\end{equation}

\begin{equation}
0=-\kappa ^p\Phi _{22},  \label{e4.5}
\end{equation}

\subsection{Calculation of $\bar \protect\sigma ^p$ and $\protect\lambda ^p$.%
}

From equation \ref{e4.1} and the fact that all the background quantities
here are real we find that

\begin{equation}
\bar{\sigma}^{p}=\left( \frac{\Delta -4\gamma +\mu }{3\Psi _{2}}\right) 
\overline{\Psi }_{0}^{\;p}.  \label{e4.6}
\end{equation}
Further, using equation \ref{e4.6} on equation \ref{e4.4} gives

\begin{equation}
\lambda ^{p}=\left( \frac{\rho -D}{3\Psi _{2}}\right) \Psi _{4}^{\;p}-\Phi
_{22}\left( \frac{\Delta -4\gamma +\mu }{9\left( \Psi _{2}\right) ^{2}}%
\right) \overline{\Psi }_{0}^{\,\,p}.  \label{e4.7}
\end{equation}

We note that the perturbed spin coefficients $\lambda ^{p}$ and $\bar{\sigma}%
^{p}$ display a definite dependence on $\Psi _{4}^{\;p}$ and/or $\overline{%
\Psi }_{0}^{\;p}$. In the rest of this section we shall derive expressions
for the remaining perturbed quantities.

\subsection{The perturbation matrix for the basis vectors.}

In order to determine the perturbed quantities $\tau ^{p}$, $\alpha ^{p}$, $%
\beta ^{p}$ and $\delta ^{p}$ and their relations , it is necessary to study
the effects of the perturbations on the basis vectors, $\left( l^{\mu
},\;n^{\mu },\;m^{\mu },\;\bar{m}^{\mu }\right) $. For compactness, it is
convenient to introduce the following index notation:

\begin{equation}
l^{1}=l^{\mu },\;l^{2}=n^{\mu },\;l^{3}=m^{\mu },\;l^{4}=\bar{m}^{\mu }.
\label{C47.1}
\end{equation}

We can write \cite{c9} the perturbations $l^{\left( p\right) i},\;\left(
i=1,2,3,4\right) $ in the vectors, as linear combinations of the unperturbed
basis vectors $l^i$. Thus

\begin{equation}
l^{\left( p\right) i}=P_{\,j}^i\,l^j,  \label{C48}
\end{equation}
where the $P_{\,j}^i$ are elements of a matrix $\mathbf{P}$ that describes,
completely, the perturbations in the basis vectors. Explicitly,

\begin{equation}
\mathbf{P}= 
\begin{array}{c}
\left[ 
\begin{array}{cccc}
P_{\,1}^1 & P_{\,2}^1 & P_{\,3}^1 & P_{\,4}^1 \\ 
P_{\,\,1}^2 & P_{\,2}^2 & P_{\,3}^2 & P_{\,4}^2 \\ 
P_{\,1}^3 & P_{\,2}^3 & P_{\,3}^3 & P_{\,4}^3 \\ 
P_{\,1}^4 & P_{\,2}^4 & P_{\,3}^4 & P_{\,4}^4
\end{array}
\right] .
\end{array}
\label{C49}
\end{equation}

The $l^{1}$ and $l^{2}$ are real while the $l^{3}$ and $l^{4}$ are complex
conjugates. It follows, then, that the matrix elements $P_{\,1}^{1}$, $%
P_{\,2}^{1}$, $P_{\,\,1}^{2}$ and $P_{\,2}^{2}$ are real while the remaining
elements of $\mathbf{P}$ are complex. Moreover, the elements in which the
indices $3$ and $4$ replace one another, are complex conjugates. For
example, $P_{\,3}^{2}=$ $\overline{\left( P_{\,4}^{2}\right) }$

\subsection{Perturbations in the Angular functions, $\protect\delta ^{p}$,$\;%
\protect\tau ^{p}$,$\;\protect\alpha ^{p}$, and $\protect\beta ^{p}$.}

The perturbations in the directional derivative $\bar{\delta}^{p}$, are
given from equations \ref{e1.7} and \ref{C48}, by

\begin{equation}
\bar \delta ^p=\left( l^4\nabla _4\right) ^p.  \label{C50}
\end{equation}
But from equations \ref{C48} and \ref{C49} we see that

\begin{equation}
l^{p\left( 4\right)
}=P_{\,j}^4\,l\,^j=P_{\,1}^4l^1+P_{\,2}^4l^2+P_{\,3}^4l^3+P_{\,4}^4l^4.
\label{C51}
\end{equation}
Using equation \ref{C51} on Equation \ref{C50} shows that

\begin{equation}
\bar \delta ^p=\left( l^4\nabla _4\right) ^p=P_{\,1}^4D+P_{\,2}^4\Delta
+P_{\,3}^4\delta +P_{\,4}^4\bar \delta .  \label{C52}
\end{equation}
Thus, if we operate with $\bar \delta ^p$ on the background $\Psi _2\,$we get

\begin{equation}
\bar{\delta}^{p}\Psi _{2}=P_{\,1}^{4}D\Psi _{2}+P_{\,2}^{4}\Delta \Psi
_{2}+P_{\,3}^{4}\delta \Psi _{2}+P_{\,4}^{4}\bar{\delta}\Psi _{2}.
\label{C53}
\end{equation}
Now, in the background

\begin{equation*}
\Psi _2=\frac{-m(u)}{r^3}. 
\end{equation*}
Substituting for $\Psi _2$ in equation \ref{C53} and using definitions of
the operators in equations \ref{e1.7} we find that

\begin{equation}
\bar \delta ^p\Psi _2=P_{\,1}^4\frac{\left[ 3m\left( u\right) \right] }{r^4}%
+P_{\,2}^4\left( \frac{-\dot m}{r^3}\right) -P_{\,2}^4\left( 1-\frac{%
2m\left( u\right) }r\right) \frac{3m\left( u\right) }{r^4}.  \label{C54}
\end{equation}

Recall that in the Vaidya spacetime background, the component $\Phi _{22}=-%
\frac{\dot{m}\left( u\right) }{r^{2}}$ is non-vanishing. Using this on
equation \ref{e4.5} shows that $\kappa ^{p}$ must vanish. It follows, then
that equation \ref{e4.2} becomes,

\begin{equation}
\bar \delta ^p\Psi _2=3\bar \tau ^p\Psi _2,  \label{C55}
\end{equation}

We see immediately, that the results expressed in equation \ref{C54} will be
inconsistent with the eigenvalue equation \ref{C55} unless the $P_{\,2}^{4}$
vanish, so that 
\begin{equation}
\bar{\delta}^{p}\Psi _{2}=P_{\,1}^{4}\frac{\left[ 3m\left( u\right) \right] 
}{r^{4}}=-3P_{\,1}^{4}\left( \frac{1}{r}\right) \Psi _{2}.  \label{C56}
\end{equation}
Equations \ref{C55} and \ref{C56}, then, show that

\begin{equation}
\bar \tau ^p=-P_{\,1}^4\left( \frac 1r\right) .  \label{C57}
\end{equation}
Moreover, equation \ref{C52} along with the condition that the $P_{\,2}^4$
vanish means that whenever $\bar \delta ^p$ acts on a function with no
angular dependence its only contribution is

\begin{equation}
\bar{\delta}^{p}=P_{\,1}^{4}\frac{\partial }{\partial r}=-\bar{\tau}^{p}r%
\frac{\partial }{\partial r}.  \label{C58}
\end{equation}
We now have, in equation \ref{C58}, a general relationship between $\bar{%
\delta}^{p}$ and $\bar{\tau}^{p}$.

Next, we need to deal with $\alpha ^{p}$ and $\bar{\beta}^{p}$. It is known 
\cite{c8} that if the null vectors $l_{\mu }$ are tangent to the geodesics
and equal to a gradient field, then

\begin{equation}
\rho =\bar \rho \;\;and\;\;\tau =\bar \alpha +\beta .  \label{C58.1}
\end{equation}
These conditions are fulfilled in all Type D space-times. In particular, the
unperturbed Vaidya space-time satisfies

\begin{equation*}
\bar{\alpha}+\beta =\tau =0.\label{C50} 
\end{equation*}
Consequently, in our linear perturbation analysis we should have

\begin{equation}
\left( \alpha +\bar \beta \right) ^p=\bar \tau ^p.  \label{C59}
\end{equation}

Consider, now, the second of the source terms in equation \ref{C37} which
reads as

\begin{equation*}
\left( \bar \delta -2\beta \right) \left( \bar \delta ^p-\bar \tau ^p+2\bar
\beta ^p+2\alpha ^p\right) \Phi _{22}. 
\end{equation*}
Our results in equations \ref{C58} and \ref{C59} when applied to the above
expression show that

\begin{equation}
\left( \bar{\delta}-2\beta \right) \left( \bar{\delta}^{p}-\bar{\tau}^{p}+2%
\bar{\beta}^{p}+2\alpha ^{p}\right) \Phi _{22}=3\left( \bar{\delta}-2\beta
\right) \bar{\tau}^{p}\Phi _{22}.  \label{C61}
\end{equation}
This achieves the purpose of expressing the effects of the four perturbed
quantities $\tau ^{p}$, $\alpha ^{p}$, $\beta ^{p}$ and $\delta ^{p}$ in
terms of one of these quantities $\tau ^{p}$. The next task is to express
this perturbed quantity in terms of the fields. To this end we utilize one
of the equations from the spin coefficient set \cite{c10}. Consider the
equation

\begin{equation}
\delta \tau -\Delta \sigma =\left( \mu \sigma +\bar{\lambda}\rho \right)
+\left( \tau +\beta -\bar{\alpha}\right) \tau -\left( 3\gamma -\bar{\gamma}%
\right) \sigma -\kappa \bar{\nu}+\Phi _{02}.  \label{C62}
\end{equation}
Specializing equation \ref{C62} to the Vaidya spacetime and applying our
linear perturbations approach we obtain, on rearranging terms, an equation
whose complex conjugate is

\begin{equation}
\left( \bar \delta -2\beta \right) \bar \tau ^p=\left( \Delta +\mu -2\gamma
\right) \bar \sigma ^p+\rho \lambda ^p.  \label{C63}
\end{equation}

Equation \ref{C63} when substituted into equation \ref{C61} gives (recall $%
\bar \delta $ does not operate on $\Phi _{22}$)

\begin{equation}
\left( \bar \delta -2\beta \right) \left( \bar \delta ^p-\bar \tau ^p+2\bar
\beta ^p+2\alpha ^p\right) \Phi _{22}=3\left[ \left( \Delta +\mu -2\gamma
\right) \bar \sigma ^p+\rho \lambda ^p\right] \Phi _{22}.  \label{C64}
\end{equation}

\subsection{The final field perturbation equations.}

We are now in position to apply the results of our discussions in this
section to the source terms of equation \ref{C35} as given by equation \ref
{C37}. Using equation \ref{C64} on equation \ref{C37} and rewriting the
sources $Q_{4}^{\prime }$ as $Q^{\prime }$ we obtain,

\begin{equation}
Q^{\prime }=\left[ \left( \Delta +2\gamma +5\mu \right) \bar{\sigma}%
^{p}-3\left( \Delta +\mu -2\gamma \right) \bar{\sigma}^{p}-3\rho \lambda
^{p}+\lambda ^{p}\left( D+\rho \right) \right] \Phi _{22}.  \label{C66}
\end{equation}
We see that in the above equation the terms in $\lambda ^{p}$ cancel
yielding the simpler result,

\begin{equation}
Q^{\prime }=-2\left( \Delta -4\gamma -\mu \right) \bar \sigma ^p\Phi _{22}.
\label{C67}
\end{equation}
Substituting for $\bar \sigma ^p$ from equation \ref{e4.6} we find that

\begin{equation}
Q^{\prime }=-2\left( \Delta -4\gamma -\mu \right) \left[ \Phi _{22}\left( 
\frac{\Delta -4\gamma +\mu }{3\Psi _{2}}\right) \overline{\Psi }_{0}^{p}%
\right] .  \label{C68}
\end{equation}
Equation \ref{C68} forms the result of our analysis in this section. All the
perturbations in the sources have now been expressed in terms of the
perturbed field $\overline{\Psi }_{0}^{p}$ only. The working field equations
(see \ref{C35} and \ref{C36}) have become

\begin{equation}
\left[ \left( \Delta +2\gamma +5\mu \right) \left( D-\rho \right) -\left(
\bar \delta -2\beta \right) \left( \delta +4\beta \right) -3\Psi _2\right]
\Psi _4^{\;p}=Q^{\prime },  \label{C69}
\end{equation}
and

\begin{equation}
\left[ \left( D-5\rho \right) \left( \Delta -4\gamma +\mu \right) -\left(
\delta -2\beta \right) \left( \bar \delta +4\beta \right) -3\Psi _2\right]
\Psi _0^{\;p}=0,  \label{C70}
\end{equation}
where, now, $Q^{\prime }$ is given by equation \ref{C68}.

Equations \ref{C69}, \ref{C70} along with \ref{C68} form the main result of
our perturbation analysis. These equations give the essential features of a
gravitationally perturbed Vaidya space time. All the non-trivial
perturbations are sufficiently described by two tetrad scalar components of
the Weyl tensor, $\Psi _{0}^{\;p}$ and $\Psi _{4}^{\;p}$, which components
represent the extreme helicity states of the gravitational field.

We now can rewrite the equations in a form that reveals the dependence of
the fields on the physical variables of spacetime. Thus, using equations \ref
{e1.7}, \ref{e1.71}, \ref{e1.9} and \ref{e1.11} on \ref{C70}, \ref{C69}, and 
\ref{C68} respectively, we find that

\ \ 

\begin{equation*}
\left[ \frac{\partial ^{2}}{\partial \,r\,\partial \,u}+\frac{5}{r}\frac{%
\partial }{\partial \,u}-\frac{1}{2}\left( 1-\frac{2m\left( u\right) }{r}%
\right) \frac{\partial ^{2}}{\partial \,r^{2}}-\frac{3}{r}\left( 1-\frac{%
m\left( u\right) }{r}\right) \frac{\partial }{\partial \,r}-\frac{2}{r^{2}}%
\right] \Psi _{0}^{\;p}+ 
\end{equation*}

\begin{equation}
\frac{1}{2\,r^{2}}\left[ \frac{\partial ^{2}}{\partial \,\theta ^{2}}+\cot
\theta \,\frac{\partial }{\partial \,\theta }-2\left( \csc ^{2}\theta +\cot
^{2}\theta \right) +\csc ^{2}\theta \frac{\partial }{\partial \,\varphi ^{2}}%
+4i\csc \theta \,\cot \theta \frac{\partial }{\partial \,\varphi }\right]
\Psi _{0}^{\;p}=0  \label{C71}
\end{equation}
and

\begin{equation*}
\left[ \frac{\partial ^2}{\partial \,u\,\partial r}+\frac 1r\frac \partial
{\partial \,u}-\frac 12\left( 1-\frac{2m\left( u\right) }r\right) \frac{%
\partial ^2}{\partial \,r^2}-\frac 1r\left( 3-\frac{7m\left( u\right) }%
r\right) \frac \partial {\partial \,r}-\frac 2{r^2}\left( 1-\frac{4m\left(
u\right) }r\right) \right] \Psi _4^{\;p} 
\end{equation*}

\begin{equation}
-\frac 1{2\,r^2}\left[ \frac{\partial ^2}{\partial \,\theta ^2}-\cot \theta
\,\frac \partial {\partial \,\theta }-2\left( \csc ^2\theta +\cot ^2\theta
\right) +\csc ^2\theta \,\frac{\partial ^2}{\partial \,\varphi ^2}-4i\csc
\theta \,\cot \theta \frac \partial {\partial \,\varphi }\right] \Psi
_4^{\;p}=Q^{\prime }  \label{C72}
\end{equation}
where now

\begin{equation*}
Q^{\prime }=-2\left[ \frac \partial {\partial \,u}-\frac 12\left( 1-\frac{%
2m\left( u\right) }r\right) \frac \partial {\partial \,r}-\frac{3m\left(
u\right) }{r^2}+\frac 1{2r}\right] \star 
\end{equation*}

\begin{equation}
\left\{ \left( \frac{\dot{m}\left( u\right) }{3m\left( u\right) }r\right) %
\left[ \frac{\partial }{\partial \,u}-\frac{1}{2}\left( 1-\frac{m\left(
u\right) }{r}\right) \frac{\partial }{\partial \,r}-\frac{m\left( u\right) }{%
r}-\frac{1}{2r}\right] \overline{\Psi }_{0}^{p}\right\} ,  \label{C73}
\end{equation}

\section{Separation of Variables}

In this section we seek to separate the equations that were derived in the
previous section. This separation of variable is effected in two phases. In
Phase I we deal with the angular variables while in Phase II we deal with
the retarded time and radial variables. We shall, for now, concentrate on
the homogeneous parts of the equations. The contribution due to the source
terms can always be constructed later once a solution for $\Psi _{0}^{\;p}$
has been obtained. Incidentally, one notices (see equation \ref{C73}) that
the luminosity-mass ratio $\frac{L}{3m\left( u\right) }$, $\left( L=-\dot{m}%
\right) $ which scales the source term will almost always be vanishingly
small since for most radiating objects the mass being radiated at any given
time is much smaller than the rest of the body mass.

\subsection{Phase I:\ The Spin-weighted Angular functions.}

We suppose that the gravitational fields entering the spherically symmetric
background spacetime are plane waves so that the problem has azimuthal
symmetry. With this, we then assume that the field equations are separable
in the angular variables admitting solutions of the form

\begin{equation}
\Psi _{i=\left( 0,4\right) }\left( u,r,\theta ,\varphi \right) =\phi
_{i=\left( 0,4\right) }\left( u,r,\theta \right) \,e^{im\varphi
}=R_{p=\left( \pm 2\right) }\left( u,r\right) \;S_{p=\left( \pm 2\right)
}\left( \theta \right) \;e.  \label{D6}
\end{equation}
Here the subscript $p$ is used to identify a particular spin-s field
component by the spin weight. For our purposes, the spin weight $p$ only
takes on the extreme values of $\pm \,s$ corresponding to the extreme
helicity states of the field. Explicitly, $\Psi _{0}$ has a spin weight of $%
2 $ while $\Psi _{4}$ has a spin weight of $-2$. Note, to avoid confusion in
notation, here and henceforth we discard the superscript $p$ previously used
to identify the perturbed quantities.

Substituting \ref{D6} in the field equations \ref{C71} and \ref{C72} yields
the following general equation in the angular variables:

\begin{equation}
\frac{1}{\sin \theta \;}\frac{d}{d\,\theta }\left( \sin \,\theta \,\frac{d}{%
d\,\theta }\right) +\left( p-p^{2}\cot \,\theta -\frac{2mp\cos \theta }{\sin
^{2}\,\theta }-\frac{m^{2}}{\sin ^{2}\,\theta }-K\right) S_{p}\left( \theta
\right) \,e^{im\varphi }=0.  \label{D10}
\end{equation}
This equation along with boundary conditions of regularity at $\theta =0$
and $\theta =\pi $ constitute a Sturm-Liouville eigenvalue problem for the
separation constant $K=\,_{p}K_{l}^{m}$. For fixed\ $p$ and $m$ values, the
eigenvalues can be labelled by $l.$ The smallest eigenvalue has $l=\max
\left( p,\left| m\right| \right) $. For each\ \thinspace $p$ and $m$ the
eigenfunctions $_{p}S_{l}^{m}\left( \theta \right) $ are complete and
orthogonal on the interval $0\leq \theta $ $\leq \pi $, as required by the
Sturm Liouville theory. In our case, where the background is non-rotating
the eigenfunctions are, the well known \cite{c14}, spin-weighted spherical
harmonics:

\begin{equation}
_pY_l^m\left( \theta ,\varphi \right) =_{\;}\,_pS_l^m\left( \theta \right)
\,e^{im\varphi }  \label{D11}
\end{equation}
and the separation constant $K$ is found to be given by

\begin{equation}
K=\;_pK_l=\left( l-p\right) \left( l+p+1\right)  \label{D12}
\end{equation}

\subsection{Phase II: The radial-null equations}

The separation of variables effected in the last section leaves us with two
equations for the functions $R_{+2}$ and $R_{-2}$. These functions are
coefficients of $_2Y_l^m\left( \theta ,\varphi \right) $ and $%
_{-2}Y_l^m\left( \theta ,\varphi \right) $, respectively, in the spin-2
fields $\Psi _0$ and $\Psi _2$ and are each dependent on $u$ and $r$ only.
On substituting equation \ref{D11} into equations \ref{C71} to \ref{C73} one
finds that $R_{+2}$ satisfies

\begin{equation*}
\left[ \frac{\partial ^2\,R_2\left( u,r\right) }{\partial r\,\partial u}%
+\frac 5r\frac{\partial R_2\left( u,r\right) }{\partial u}-\frac 12\left( 1-%
\frac{2m\left( u\right) }r\right) \frac{\partial ^2R_2}{\partial r}\right] 
\end{equation*}

\begin{equation}
-\left[ \frac 3r\left( 1-\frac{m\left( u\right) }r\right) \frac{\partial
R_2\left( u,r\right) }{\partial \,r}-\frac{\left( _2K_l-4\right) }{2r^2}%
R_2\left( u,r\right) \right] =0.  \label{D13}
\end{equation}

and $R_{-2}$ satisfies 
\begin{equation*}
\left[ \frac{\partial ^{2}R_{-2}}{\partial u\partial r\,}+\frac{1}{r}\frac{%
\partial R_{-2}}{\partial u}-\frac{1}{2}\left( 1-\frac{2m\left( u\right) }{r}%
\right) \frac{\partial ^{2}R_{-2}}{\partial r^{2}}-\left( \frac{3}{r}-\frac{%
7m\left( u\right) }{r^{2}}\right) \frac{\partial R_{-2}}{\partial \,r^{2}}%
\right] 
\end{equation*}

\begin{equation}
+\left( \frac 1{2r^2}\left( _{-2}K-4\right) +\frac{8m\left( u\right) }{r^3}%
\right) R_{-2}\left( u,r\right) =0  \label{D14}
\end{equation}

We deal with equations \ref{D13} and \ref{D14} separately. First, we shall
seek to separate equation \ref{D13} in $R_2\left( u,r\right) $. By adopting
a change of variables we show that the equation is separable for a specific
choice of mass function. Thereafter, we shall apply this approach to
equation \ref{D14}.

\subsubsection{Change of variables.}

Equation \ref{D13}, as it stands, is not separable. We shall, therefore,
find it convenient to introduce the following change of variables: let us set

\begin{equation}
\tau =\frac 1u\;\;\;\;and\;\;\;\xi =\frac{2m\left( u\right) }r.  \label{D15}
\end{equation}
Then it is seen that\ 
\begin{equation*}
\frac \partial {\partial u}=-\tau ^2\,\frac \partial {\partial \tau }+\frac{%
\dot m}{m\left( \tau \right) }\xi \,\frac \partial {\partial \xi }, 
\end{equation*}

\begin{equation}
\frac \partial {\partial r}=-\frac{\xi ^2}{2m\left( \tau \right) }\,\frac
\partial {\partial \xi },  \label{D16}
\end{equation}
and 
\begin{equation*}
\frac{\partial ^2}{\partial r^2}=\frac{\xi ^3}{2\left[ m\left( \tau \right) %
\right] ^2}\,\frac \partial {\partial \xi }+\frac{\xi ^4}{4\left[ m\left(
\tau \right) \right] ^2}\,\frac{\partial ^2}{\partial \xi ^2}, 
\end{equation*}
where, now, $m$ is a function of $\tau $ but $\dot m$ still means $\frac{dm}{%
d\,u}$.

\subsection{Equation for R$_{+2}$.}

On substituting equations \ref{D15} and \ref{D16} into \ref{D13} and
rearranging we find that

\begin{equation*}
\left( -5\xi \,\frac{\partial R_2}{\partial \tau }+\xi ^2\frac{\,\partial
^2\,R_2}{\partial \xi \;\partial \tau }\right) 4\,\tau ^2m\left( \tau
\right) -4\dot m\left( \xi ^3\,\frac{\partial ^2R_2}{\partial \xi ^2}-4\xi ^2%
\frac{\partial R_2}{\partial \xi }\right) +\left( \xi ^5-\xi ^4\right) \frac{%
\partial ^2R_2}{\partial \xi ^2} 
\end{equation*}

\begin{equation}
-\left( \xi ^{4}-4\xi ^{3}\right) \frac{\partial R_{2}}{\partial \xi }%
+\varepsilon _{l}\,\xi ^{2\,}R_{2}\left( \tau ,\xi \right) =0  \label{D17}
\end{equation}
where we have set 
\begin{equation}
\varepsilon _{l}=\left( l-p\right) \left( l+p+1\right) -4=\left( l-2\right)
\left( l+3\right) -4  \label{D17.1}
\end{equation}

We would like, as an example, to apply our analysis on an evaporating
blackhole. In general, the equation \ref{D17} is not separable for an
arbitrary mass function $m\left( u\right) $. However it can be shown to
separate for one particular model of such a radiating blackhole.

\ 

\paragraph{Vaidya model for a linearly radiating blackhole.}

\ \ 

In the Vaidya model of a radiating blackhole \cite{c14.1}, the spacetime is,
initially, Minkowski flat for $u<0$. Then at $u=0$ an imploding $\delta -$%
function-like null fluid with a total positive mass $M$ forms a blackhole.
Hereafter, $0<u<u_{0}$ negative energy null fluid then falls into the
blackhole evaporating the latter in the process. One known consequence \cite
{c14.1} is that the spacetime violates the weak energy condition. Eventually
the blackhole vanishes so that for $u\geq u_{0}$ the spacetime becomes
Minkowski flat again.

One of the popular models of radiating black holes is the so-called
self-similar model originally developed by Hiscock, \cite{c14.1}. Popular,
because from it one can construct the quantum energy stress tensor for the
entire spacetime. The model has been extensively used lately, (see for
example \cite{c14.2} and \cite{c14.3}\textit{). }In this model the mass is a
linear function of the retarded time coordinate $u$.

We shall show, presently, that for the Hiscock linear mass function ansatz
the above equation is separable.

Suppose 
\begin{equation}
m\left( u\right) =\left\{ 
\begin{tabular}{l}
$0,\;u<0\;$ \\ 
$M_{0}\left( 1-\lambda \,u\right) ,\;\frac{1}{\lambda }>u>0$ \\ 
$0,\;u>\frac{1}{\lambda }$%
\end{tabular}
\right\}  \label{D18}
\end{equation}
so that 
\begin{equation}
\dot{m}=-\lambda M_{0}  \label{D18.1}
\end{equation}
where $m_{0}$ is the initial mass at $u=0$ and $\lambda $ is some positive
parameter $0<\lambda <\frac{1}{u}$ that scales the radiation rate.

We shall find it convenient to institute a change of variable $u=v_{0}+v$
where $v_{0}$ is some fixed value of $u$. 
\begin{equation}
m\left( v\right) =\left\{ 
\begin{tabular}{l}
$m_{0}\left( 1+\lambda v\right) ,\;-v_{0}<v<0$ \\ 
$m_{0},\;\;\;\;\;\;\;\;\;\;\;v=0$ \\ 
$m_{0}\left[ 1-\lambda \,v\right] ,\;\frac{1}{\lambda }>v>0$ \\ 
$0,\;v>\frac{1}{\lambda }$%
\end{tabular}
\right\}  \label{D19}
\end{equation}
This seemingly trivial change is important for the following reasons. In our
problem we would like to discuss the behavior of the gravitational fields in
a radiating blackhole background. However as we noted above just before $u=0$
there is no black hole and yet we need the ingoing fields to have been
moving in a non-minkowski background. This change, therefore, makes it
physically possible for us to introduce the external gravitational fields
into a spacetime that already contains the black hole. Mathematically the
change makes it possible, as we find out soon, to construct complete
solutions that include a description of ingoing fields.

Substituting equation \ref{D18} into \ref{D17} and making use of equation 
\ref{D15} we find that

\begin{equation*}
\left( -5\xi \frac{\,\partial R_{2}}{\partial \tau }+\xi ^{2}\,\frac{%
\partial ^{2}R_{2}}{\partial \xi \,\partial \tau }\right) 4\,\tau
^{2}m\left( \tau \right) +,\left( \xi ^{5}-\xi ^{4}+4\lambda m_{0}\,\xi
^{3}\right) \frac{\partial ^{2}R_{2}}{\partial \xi ^{2}} 
\end{equation*}

\begin{equation*}
\left( -5\xi \frac{\,\partial R_2}{\partial \tau }+\xi ^2\,\frac{\partial
^2R_2}{\partial \xi \,\partial \tau }\right) 4\,\tau ^2m\left( \tau \right)
+\left( \xi ^5-\xi ^4+4\lambda m_0\,\xi ^3\right) \frac{\partial ^2R_2}{%
\partial \xi ^2} 
\end{equation*}

\begin{equation}
-\left( \xi ^4-4\xi ^3+16\lambda \,m_0\xi ^2\right) \frac{\partial R_2}{%
\partial \xi }+\varepsilon _l\,\xi ^2\,R_2\left( \tau ,\xi \right) =0
\label{D20}
\end{equation}

\subsubsection{Separability.}

We now ask whether the equation \ref{D20} is separable (in the variables $%
\tau $ and $\xi $) admitting a solution of the form 
\begin{equation}
R_{2}\left( \tau ,\xi \right) =\;X_{2}\left( \xi \right) \,Y_{2}\left( \tau
\right)  \label{D20.1}
\end{equation}
where $X_{2}$ and $Y_{2}$ are each functions of one variable, only.
Substituting for $R_{2}$ in equation \ref{D20} yields two ordinary
differential equations: a first order differential equation for $Y_{2}\left(
\tau \right) $, 
\begin{equation}
m\tau ^{2}\left( \frac{1}{Y\left( \tau \right) }\,\frac{dY_{2}}{d\tau }%
\right) =-\alpha  \label{D22}
\end{equation}
and a second order ordinary differential equation for $X_{2}\left( \xi
\right) $,

\begin{equation}
\xi ^{2}\left( \xi ^{2}-\xi +4\lambda m_{0}\right) \frac{d^{2}X_{2}}{d\,\xi
^{2}}-\xi \left[ \xi ^{2}-4\xi +4\alpha +16\lambda m_{0}\right] \frac{dX_{2}%
}{d\xi }+\left( \varepsilon _{l}\,\xi +20\alpha \right) X_{2}\left( \xi
\right) =0.  \label{D23}
\end{equation}
Here, $\alpha $ is an arbitrary separation constant. Its characteristics are
discussed in the next section.

\subsection{The function $R_{-2}$}

Following the same approach as above we find that the equation for the
function $R_{-2}$ separates into two ordinary differential equations:

\begin{equation}
m\left( \tau \right) \tau ^2\frac 1{Y_{-2}}\frac{dY_{-2}}{d\tau }=-\gamma ,
\label{D30}
\end{equation}
and 
\begin{equation*}
\xi ^2\left( \xi ^2-\xi +4m_0\lambda \right) \frac{d^2X_{-2}}{d\,\xi ^2}-\xi %
\left[ 5\xi ^2-4\xi +4\gamma \right] \frac{dX_{-2}}{d\xi } 
\end{equation*}

\begin{equation}
+\left( \xi ^{2}+2\varepsilon _{l}\,\xi +4\gamma \right) X_{-2}=0,
\label{D32}
\end{equation}
The general characteristics of the separation constant $\gamma $ are not
different from those of $\alpha $ and discussed in the next section.

\section{Solutions}

We have achieved the separation of the original equations \ref{D13} and \ref
{D14} respectively, into \ref{D22}, \ref{D23} and \ref{D30} and \ref{D32}
for the particular case of a linear mass function. In the following sections
we shall seek to solve these resulting ordinary differential equations and
to discuss the solutions in an attempt to draw some physical information
from them.\ \ \ \ \ \ 

\subsection{The functions $Y_{p}\left( \protect\tau \right) $}

The first order differential equations for $Y_{2}\left( \tau \right) $ and $%
Y_{-2}\left( \tau \right) $ above can be integrated immediately. Thus from
equation \ref{D22} we find that 
\begin{equation}
Y_{2}\left( v\right) =\exp \left( -\alpha \int_{0}^{u}\frac{dv}{m_{0}\left(
1-\lambda v\right) }\right) =e^{^{\frac{\alpha }{\lambda m_{0}\,}\ln \left(
1-\lambda v\right) }}.  \label{A8}
\end{equation}
Now $0\leq \lambda <1$ and in fact for most radiating bodies $\lambda \ll 1$%
. This allows us to expand the logarithmic expression $\ln \left( 1-\lambda
u\right) $ in the solution so that

\begin{equation}
Y_{2}\left( v\right) =e^{-\Omega v}e^{-\Omega \overset{\infty }{%
\sum\limits_{n=1}}\frac{1}{n+1}\lambda ^{n}v^{n+1}},  \label{A9}
\end{equation}
where the separation constant now takes the form $\Omega $ in which we
absorb the Schwarzschild mass $m_{0}$

\begin{equation}
\alpha =m_0\Omega .  \label{A10}
\end{equation}
Consider, now, the case in which the background is not radiating. It is
clear either from the solution above or from the original differential
equation that for such a case the solution reduces to

\begin{equation}
Y_{2}\left( v_{s}\right) =e^{-\Omega _{s}v_{s}},  \label{A10.1}
\end{equation}
where the subscript $s$ indicates quantities associated with the
Schwarzschild geometry. One notices that in such a static background the
quantity $Y_{2}\left( v_{s}\right) $ above constitutes the only time
dependent part of the $\Psi _{0}$ field. It follows then that to be
consistent with the known \cite{c3} solutions we should require

\begin{equation}
\Omega _s=i\omega ,  \label{A10.2}
\end{equation}
where $\omega $ is the frequency of the gravitational waves. This suggests
that in the case of the radiating background we should expect the parameter $%
\Omega $ to be a complex function of $\lambda $ and $\omega $ such that 
\begin{equation}
\lim\limits_{\lambda \rightarrow 0}\Omega \left( \lambda ,\omega \right)
=i\omega .  \label{A12}
\end{equation}

The integration of the differential equation for $Y_{-2}\left( v\right) $
follows the same trend and we find that 
\begin{equation}
Y_{-2}\left( v\right) =e^{-\Gamma v}e^{-\Gamma \overset{\infty }{%
\sum\limits_{n=1}}\frac{1}{n+1}\lambda ^{n}v^{n+1}},  \label{A13}
\end{equation}
where

\begin{equation}
\gamma =m_0\Gamma ,  \label{A14}
\end{equation}
$\Gamma $ being a complex function of $\lambda $ and $\omega $ such that 
\begin{equation}
\lim\limits_{\lambda \rightarrow 0}\Gamma =i\omega  \label{A16}
\end{equation}

As would be expected from the theory of deferential equations the separation
constants $\Omega $ and $\Gamma $ can not have unique values. The individual
solutions we obtain will therefore be representatives of classes of
solutions. The range of these solutions is described in terms of the
frequency spectrum of the gravitational field which, in our classical
treatment, takes on continuous values. It will, later, be shown that by
using certain conditions on the solutions the functional form of these
separation constants can be more rigidly fixed.

\subsection{The functions $X_{p}\left( \protect\xi \right) $}

Following the integrations of the first order differential equations for $%
Y_p $ we are now left with the two equations for $X_2$ and $X_{-2}$ to
solve. These are respectively,

\begin{equation}
\xi ^{2}\left( \xi ^{2}-\xi +4\lambda m_{0}\right) \frac{d^{2}X_{2}}{d\,\xi
^{2}}-\xi \left[ \xi ^{2}-4\xi -4m_{0}\left( 4\lambda -\Omega \right) \right]
\frac{dX_{2}}{d\xi }+\left( \varepsilon _{l}\,\xi -20m_{0}\Omega \right)
X_{2}\left( \xi \right) =0  \label{A17}
\end{equation}
and 
\begin{equation*}
\xi ^{2}\left( \xi ^{2}-\xi +4m_{0}\lambda \right) \frac{d^{2}X_{-2}}{d\,\xi
^{2}}-\xi \left[ 5\xi ^{2}-4\xi +4m_{0}\Gamma \right] \frac{dX_{-2}}{d\xi } 
\end{equation*}

\begin{equation}
+\left( \xi ^{2}+2\varepsilon _{l}\,\xi +4m_{0}\Gamma \right) X_{-2}=0.
\label{A18}
\end{equation}
It is clear that at $\xi =0$ (or $r=\infty $) both the equations above have
regular singularities \cite{c14.4}. This encourages us to seek for analytic
solutions. Such solutions at $\xi =0\,\left( r=\infty \right) $ should be
useful in discussing the asymptotic fall-offs of the fields and the question
of energy flux.

\subsubsection{The peeling behavior}

Our initial goal is to develop asymptotic solutions for the functions $%
X_{2}\left( \xi \right) $ and $X_{-2}\left( \xi \right) $. Consider a zero
rest mass spin-s field $\psi _{p}$ in a helicity state $p$. According to the
peeling theorem by Roger Penrose \cite{c8}, the quantities $r^{(s+p+1)}\psi
_{p}$ and $r^{(s-p+1)}\psi _{p}$ have a limit at null-infinity. In the case
of gravitational fields we expect the outgoing components of the solutions
to fall off as 
\begin{equation}
\psi _{\left( p=\pm 2\right) }\sim \frac{1}{r^{(s+p+1)}}=\frac{1}{r^{\left(
2\pm 2+1\right) }},  \label{A24}
\end{equation}
while the ingoing solutions should fall off as 
\begin{equation}
\psi _{\left( p=\pm 2\right) }\sim \frac{1}{r^{(s-p+1)}}=\frac{1}{r^{\left(
2\mp 2+1\right) }}.  \label{A25}
\end{equation}
It is necessary, therefore, that the solutions to our differential equations
display the above asymptotic behavior. This, indeed, will be one of the
tests for their validity.

\subsubsection{The Indicial Equations.}

It has been pointed out that at $\xi =0$ we have a regular singularity in
both equations \ref{A17} and \ref{A18} . Therefore it seems natural to
attempt developing solutions about this point. Such solutions will be valid
at far distances from the black hole. This class of solutions at such
distances is useful if one is to engage, as we shall later, in a meaningful
discussion of the gravitational energy flux.

Let us assume that equation \ref{A17} admits, as a solution, a series
expansion about $\xi =0$ of the form

\begin{equation}
X_2\left( \xi \right) =\overset{\infty }{\sum\limits_{n=0}}a_n\xi ^{n+k},
\label{A26}
\end{equation}
where $k$ is some value to be determined. Using equation \ref{A26} in
equation \ref{A17} gives

\begin{equation*}
\overset{\infty }{\sum\limits_{n=0}}a_n\left( n+k\right) \left( n+k-2\right)
\xi ^{n+k+2} 
\end{equation*}

\begin{equation*}
-\overset{\infty }{\sum\limits_{n=0}}a_n\left[ \left( n+k\right) \left(
n+k-5\right) -\varepsilon _l\right] \xi ^{n+k+1} 
\end{equation*}

\begin{equation}
\overset{\infty }{+\sum\limits_{n=0}4}a_n\left\{ \left( n+k\right) \left[
\left( n+k-5\right) \lambda m_0-m_0\Omega \right] +5m_0\Omega \right\} \xi
^{n+k}=0  \label{A27}
\end{equation}
For $n=0,\;a_0\neq 0$ we get the indicial equation

\begin{equation}
k\left[ \left( k-5\right) \lambda -\Omega \right] +5\Omega =0.  \label{A28}
\end{equation}
which has two distinct roots,

\begin{equation}
k=\left( 5,\frac \Omega \lambda \right) .  \label{A29}
\end{equation}

Similarly, for $X_{-2}\left( \xi \right) $, we can assume a solution of the
form

\begin{equation}
X_{-2}\left( \xi \right) =\overset{\infty }{\sum\limits_{n=0}}b_n\xi ^{n+s},
\label{A30}
\end{equation}
where, again $s$ is some value to be determined. Substituting equation \ref
{A30} into equation \ref{A18}, we find that

\begin{equation*}
\overset{\infty }{\sum\limits_{n=2}}b_{n-2}\left[ \left( n+s-2\right) \left(
n+s-8\right) +1\right] \xi ^{n+s} 
\end{equation*}

\begin{equation*}
-\overset{\infty }{\sum\limits_{n=1}}b_{n-1}\left[ \left( n+s-1\right)
\left( n+s-6\right) +2\varepsilon _l\right] \xi ^{n+s} 
\end{equation*}

\begin{equation}
+4m_0\overset{\infty }{\sum\limits_{n=0}}b_n\left[ \lambda \left( n+s\right)
\left( n+s-1\right) -\Gamma \left( n+s\right) +\Gamma \right] \xi ^{n+s}=0
\label{A31}
\end{equation}

From the equation above and the condition that $b_0\neq 0$ we obtain, for $%
n=0$, the indicial equation

\begin{equation}
s^2-\left( 1+\frac \Gamma \lambda \right) s+\frac \Gamma \lambda =0
\label{A32}
\end{equation}
whose roots are

\begin{equation}
s=\left( 1,\frac \Gamma \lambda \right) .  \label{A33}
\end{equation}
Equations \ref{A29} and \ref{A33} (and the fact that $\Omega $ and $\Gamma $
are complex quantities) indicate that we can expect two linearly independent
solutions for each of the fields.

\paragraph{Asymptotic conditions}

The solutions to the indicial equations given in equations \ref{A29} and \ref
{A33} fix for us the leading terms for the functions $X_2\left( \xi \right) $
and $X_{-2}\left( \xi \right) $, respectively. Thus from equation \ref{A29}
we see that

\begin{equation}
X_{2}\left( \xi \right) \sim \xi ^{5}\text{ or }\xi ^{\frac{\Omega }{\lambda 
}},  \label{A34}
\end{equation}
and from equation \ref{A33} 
\begin{equation}
X_{-2}\left( \xi \right) \sim \xi \text{ or }\xi ^{\frac{\Gamma }{\lambda }}.
\label{A35}
\end{equation}
Both equations \ref{A34} and \ref{A35} show that the first solutions are
consistent with the peeling theorem and can, in fact be recognized as
outgoing fields (recall $\xi =\frac{2m\left( v\right) }{r}=\frac{%
2m_{0}\left( 1-\lambda v\right) }{r}$).

On the other hand the second solutions are scaled by the quantities $\Omega $
and $\Gamma $, respectively. These are the same arbitrary separation
constants which, in the last section, we showed to be complex. Since,
physically, our solutions represent gravitational fields these constants
must now be chosen to conform with the known boundary values for such
ingoing waves. Consequently, in order to satisfy the peeling theorem, it is
clear that we must have $Re\Omega \sim \lambda $, so that $X_{2}\sim \frac{1%
}{r}\ $and $Re\Gamma \sim 5\lambda $, so that $X_{-2}\sim \frac{1}{r^{5}}$.
Moreover, the imaginary parts of these quantities must reduce to the
limiting cases, $\lim\limits_{\lambda \rightarrow 0}\Omega =i\omega $ and $%
\lim\limits_{\lambda \rightarrow 0}\Gamma =i\omega $ as was shown to be the
case. These two conditions dictate that we set

\begin{equation}
\Omega =\lambda +i\omega ,  \label{A36}
\end{equation}
and 
\begin{equation}
\Gamma =5\lambda +i\omega ,  \label{A37}
\end{equation}
The roots to the indicial equations \ref{A29} and \ref{A33}, respectively,
now become

\begin{equation}
k=\left[ 5,\left( 1+\frac i\lambda \omega \right) \right] ,  \label{A38}
\end{equation}

and

\begin{equation}
s=\left[ 1,\left( 5+\frac i\lambda \omega \right) \right] .  \label{A38.0}
\end{equation}
So that as $\xi \rightarrow 0$,

\begin{equation}
X_{2}\left( \xi \right) \rightarrow \xi ^{5}\;or\;X_{2}\left( \xi \right)
\rightarrow \xi \xi ^{\frac{i}{\lambda }\omega },  \label{A39}
\end{equation}
and 
\begin{equation}
X_{-2}\left( \xi \right) \rightarrow \xi \;or\;X_{-2}\left( \xi \right)
\rightarrow \xi ^{5}\xi ^{\frac{i}{\lambda }\omega }.  \label{A40}
\end{equation}
The full functions $X_{2}\left( \xi \right) $ are readily obtained by
writing down recurrence relations using equations \ref{A27} and \ref{A31}.
The general solutions are, in each case, found to be linear combinations of
the outgoing component $X_{p}^{\left( out\right) }$ and the ingoing
component $X_{p}^{\left( in\right) }$.

Thus

\begin{equation}
X_{p}\left( \xi \right) =A_{p}X_{p}^{\left( out\right) }+B_{p}X_{p}^{\left(
in\right) }.  \label{A41}
\end{equation}
Here the $A_{p}$ and $B_{p}$ are arbitrary constants of integration. These
solutions are also found to converge.

\section{The asymptotic solutions and physical information.}

\subsection{Significance}

The principal aim of our study is to understand how gravitational waves are
scattered by a background radiating spacetime. In particular, we are
interested in the measurable physical results of this process, such as the
energy flux and the manner in which the waves are reflected and absorbed by
a radiating black hole. To this end we have, in the preceding discussions,
developed field equations that describe the effects of these waves on the
background spacetime. The physical quantities that we seek should, in
principle, be calculated from the solutions of these equations. As can
easily be shown, however, the series solutions obtainable are a result, in
each case, of a three term recursion relation and so contain various
coefficients that are not easy to relate. This feature of our solutions
would seem to make inconvenient, their use in calculating a number of other
physical quantities. It turns out, though, that for the features of our
interest it is sufficient to consider the form of the solutions at certain
special points. For example Chandrasekhar \cite{c9} shows that a knowledge
of the incident and reflected wave amplitudes can be deduced from the form
of the solution at null-infinity. Moreover, one can also engage in a
meaningful discussion pertaining to energy flux at these points. This means
that we need only consider the leading terms in the solutions.

\subsection{The Source Terms}

In creating the function $\Psi _{4}$ we have, so far, only considered
solutions for the homogeneous part of the original differential equation
(see equations \ref{C72} and \ref{C73}). However, the full equation for $%
\Psi _{4}$ is inhomogeneous so that the complete solution should contain a
contribution due to the sources. We recall that the source term is scaled by
the luminosity $L=-\frac{dm}{dv}=\lambda m_{0}$ which obviously vanishes as
the background radiation is switched off, $\lambda \longrightarrow 0$.
Since, in the first place, $\lambda \ll 1\Longrightarrow \frac{\dot{m}\left(
v\right) }{3m\left( v\right) }\ll 1$, we shall presently assume that at
large $r$ values the source terms do not contribute significantly to the
solution. Consequently we shall consider the asymptotic solutions from the
homogeneous equation to be a sufficient representation of the general
asymptotic solutions. With this we now write down the asymptotic form of the
entire solutions.

\subsection{The Solutions}

It was shown, in equation \ref{A9}, that for $\Psi _{0}$,

\begin{equation}
Y_{2}\left( u\right) =e^{-\frac{\Omega }{\lambda }\ln \left( 1-\lambda
v\right) }=e^{-\Omega u}e^{\overset{\infty }{-\sum\limits_{n=1}}\frac{1}{n+1}%
\lambda ^{n}v^{n+1}}.  \label{A44}
\end{equation}
For small $\lambda $ values, the expression for the logarithmic expansion
can be written to first order in $\lambda $. This gives 
\begin{equation}
Y_{2}\left( u\right) =e^{\frac{\Omega }{\lambda }\ln \left( 1-\lambda
v\right) }\simeq e^{-\Omega v}.  \label{A45}
\end{equation}
And using the conditions spelled in equation \ref{A36} to satisfy the
peeling property for the ingoing field, we find that 
\begin{equation}
Y_{2}\left( u\right) \simeq e^{-\lambda v-i\omega v}  \label{A46}
\end{equation}
Similarly, going through the same treatment for $Y_{-2}\left( u\right) $ and
applying equation \ref{A37}, we find that 
\begin{equation}
Y_{-2}\left( u\right) =e^{-5\lambda v-i\omega v}  \label{A47}
\end{equation}
The above results along with the angular solutions $_{p}Y_{l}^{m}\left(
\theta ,\varphi \right) =_{\;}\,_{p}S_{l}^{m}\left( \theta \right)
\,e^{im\varphi }$ of equation \ref{D11} can now be joined to the functions
of equations \ref{A39} and \ref{A40} (recall $\xi =\frac{2m\left( v\right) }{%
r}=\frac{2m_{0}\left( 1-\lambda v\right) }{r}$) to give the following
asymptotic solutions for the functions $\Psi _{0}$ and $\Psi _{4}$:

\ \ 
\begin{equation}
\begin{array}{lll}
& 
\begin{array}{l}
outgoing \\ 
\;
\end{array}
& 
\begin{array}{l}
ingoing \\ 
\;
\end{array}
\\ 
\begin{array}{l}
\Psi _{0}\;\;\sim \\ 
\;
\end{array}
& 
\begin{array}{l}
_{2}Y_{l}^{m}\left( \theta ,\varphi \right) \frac{\left[ 2m\left( u\right) 
\right] ^{5}}{r^{5}}e^{-\left( \frac{L}{m_{0}}\right) v}e^{-i\omega v} \\ 
\;
\end{array}
& 
\begin{array}{l}
_{2}Y_{l}^{m}\left( \theta ,\varphi \right) \frac{2m\left( u\right) }{r}%
e^{-\left( \frac{L}{m_{0}}\right) v}e^{ip}e^{-2i\omega v} \\ 
\;
\end{array}
\\ 
\Psi _{4}\;\;\sim & _{-2}Y_{l}^{m}\left( \theta ,\varphi \right) \frac{%
2m\left( u\right) }{r}e^{-5\left( \frac{L}{m_{0}}\right) v}e^{-i\omega v} & 
_{-2}Y_{l}^{m}\left( \theta ,\varphi \right) \frac{\left[ 2m\left( u\right) 
\right] ^{5}}{r^{5}}e^{-5\left( \frac{L}{m_{0}}\right) v}e^{ip}e^{-2i\omega
v}
\end{array}
\label{A48}
\end{equation}

\ \ 

Here, $p\left( r\right) =\frac{\omega }{\lambda }\ln \left( \frac{2m_{0}}{r}%
\right) $ and where for physical reasons we find it useful to express the
solutions in terms of the luminosity $L$ as given by $L=-\frac{dm}{dv}%
=\lambda m_{0}.$

\section{Conclusion}

We have obtained analytic solutions to the problem of gravitational fields
propagating in a radiating spacetime. These solutions satisfy all the known
conditions for the propagation of spin-s, zero rest mass fields. From their
asymptotic form in equations \ref{A48} it is seen that the solutions are
completely consistent with the peeling theorem of Penrose and fall off in
the manner predicted by this theorem. Moreover, one observes, further, that
as the background radiation is switched off (i.e. in the limit $\lambda
\rightarrow 0$), the theory recovers the known solutions (see for example 
\cite{c3}) for the perturbed static geometry of Schwarzschild. We consider
the passing of these two tests a validation of our analysis.

One new significant feature this analysis brings to surface is that the
solutions (see equation \ref{A48}) are scaled by factors of the form $%
e^{-\left( s+p+1\right) \left( \frac{L}{m_{0}}\right) u}$, where $s=2$ and $%
p=\pm 2$. But $\frac{L}{m_{0}}$ is positive definite. Consequently, these
factors indicate that when gravitational fields propagate in a radiating
spacetime they suffer an attenuation, and this attenuation can be
quantitatively described. The attenuation weight seems in turn to be
directly related to the spin weight of the perturbed fields. It is also
scaled by the luminosity $L$ of the background. Further, as one notices from
the solutions, the attenuation persists independent of whether the fields
are ingoing or outgoing. It is of course fair to ask whether this character
of our solutions is not, in the first place, a reflection of the mass
function that we chose. Recalling that general radiative mass function $%
m\left( v\right) $ is a monotonic decreasing function in $v$ an expansion of
the $m\left( v\right) =m_{0}-Lm_{0}-\frac{dL}{2!dv}m_{0}-....$ about $v=0$
indicates that the first order term in the luminosity would seem to make the
significant contribution. This seems to suggest that the attenuations
manifested in our solutions are independent of the manner in which the
blackhole radiates and may persist for any mass function chosen. As far as
we know this seems to be a new feature in the literature of this branch of
general relativity; one that may, indeed, have some interesting
astrophysical implications.

A persistent attenuation of this sort would seem to suggest the possibility
that energy is being dumped into the host spacetime. Over large time scales,
this could have significant implications on the evolution of such a
radiating system. This question can, however, only be resolved by a rigorous
calculation of the energy flux. For such a calculation and an extension of
this discussion see \cite{c16} We intend to follow up this issue in future
discussions.

\end{document}